\documentclass[amsmath,amssymb,aps,pre,superscriptaddress,citeautoscript,twocolumn,floatfix]{revtex4-2}
\usepackage[bookmarks=false]{hyperref}
\usepackage{graphicx}
\usepackage{dcolumn}
\usepackage{bm}
\usepackage{xcolor}

\newcommand{\ps}[1]

\def\bnabla{\mbox{\boldmath $\nabla $}}

\begin{document}

\preprint{APS/123-QED}
\title{On the nature of screening in Voorn-Overbeek type theories}
\author{Sunita Kumari}
\affiliation{School of Physical Sciences, University of Chinese Academy of Sciences, Beijing 100049, China}
\author{Shikha Dwivedi}
\affiliation{School of Physical Sciences, University of Chinese Academy of Sciences, Beijing 100049, China}
\author{Rudolf Podgornik}
\affiliation{School of Physical Sciences, University of Chinese Academy of Sciences, Beijing 100049, China}
\affiliation{Kavli Institute for Theoretical Sciences, University of Chinese Academy of Sciences, Beijing 100049, China}
\affiliation{ Institute of Physics, Chinese Academy of Sciences, Beijing 100190, China}
\affiliation{Wenzhou Institute of the University of Chinese Academy of Sciences, Wenzhou, Zhejiang 325000, China}
\altaffiliation{Department of theoretical physics, J. Stefan Institute, , 1000 Ljubljana, Slovenia and Department of Physics, Faculty of Mathematics and Physics, University of Ljubljana, 1000 Ljubljana, Slovenia}
 \email{podgornikrudolf@ucas.ac.cn}

\date{\today}

\begin{abstract}
By using a recently formulated Legendre transform approach to the thermodynamics of charged systems, we explore the general form of the screening length in the Voorn-Overbeek-type theories, that remains valid also in the cases where the entropy of the charged component(s) is not given by the ideal gas form as in the Debye-H\"uckel theory. The screening length consistent with the non-electrostatic terms in the free energy {\sl Ansatz} for the Flory-Huggins and Voorn-Overbeek type theories, derived from the local curvature properties of the Legendre transform,  has distinctly different behavior than the often invoked standard Debye screening length, though it reduces to it in some special cases.  
\end{abstract}

\maketitle


\section{Introduction}

{While it was originally formulated in the context of phase separation between charged polyions, different variants of the  Voorn-Overbeek (VO) theory  \cite{voorn1956} still represent the basic conceptual underpinning of diverse phenomena ranging from the theory of phase separation in solutions of weakly charged polyelectrolytes  \cite{Monica2004,Monica2004a}, swelling of charged gels \cite{Muthu2012,Muthu2021}, polyelectrolyte complex coacervates \cite{Srivastava, Sing2020,Wang2021}, as well as liquid-liquid phase coexistence phenomena in biology \cite{brangwynne2015, PERRY201986}.} 

Formally, the VO theory is anchored in the competition between the configurational entropy of charged polyions and electrostatic correlation attraction between them, the former being evaluated within the Flory-Huggins polymer mixing framework \cite{Teraoka}, while the latter is based on the Poisson-Boltzmann (PB) electrolyte theory in the Debye-H\"{u}ckel (DH) approximation  \cite{muthu2002}. The system is modelled as consisting of three components: water and two types of  polyions, denoted by $p^+$ and $p^-$, of charge $\pm e N_{p^{\pm}}$, and degree of polymerization $N_{p^{\pm}}$. The monomers and water molecules are all assumed to have the same molecular volume, $v = a^3$. 

The components of the VO free energy density, following the local density approximation where the inhomogeneities described only by the coordinate dependence of the densities, are then assumed to have the following forms: 

(i) The Flory-Huggins polymer solution free energy is given by \cite{Doi,Teraoka}
\begin{widetext}
\begin{eqnarray}
\frac{f_{FH}(\phi_{p_+}, \phi_{p_-})~a^3}{k_B T} = {\frac {\phi_{p_+}}{N^+}}\ln \phi_{p_+} + {\frac {\phi_{p_-}}{N^-}}\ln \phi_{p_-}  + {(1-\phi_{p_+}-\phi_{p_-})}\ln(1-\phi_{p_+}-\phi_{p_-}) + {\textstyle\frac{1}{2}} \sum_{j,k}\chi_{jk}\phi_{j}\phi_{k},
\label{e2}
\end{eqnarray}
\end{widetext}
where $f_{FH}$ is the free energy density, $k_BT$ is the thermal energy and $\phi_{\pm}$ the volume fraction of $p^\pm$ polymers. 

We included also the $\chi_{jk}$ interaction term,  describing the short range interactions of non-electrostatic nature such as the van der Waals interactions, which is also frequently included in later formulations of the theory \cite{Larson2016}.  The connection between the volume fractions $\phi_{p^{\pm}}$ and the concentrations $c_{p^{\pm}}$ is given by $ \phi_{p^{\pm}} = a^3 c_{p^{\pm}} N_{p^{\pm}}$. 

(ii) The DH dilute electrolyte correlation free energy can be obtained in different ways \cite{FALKENHAGEN19711,McQuarrie,Levin_2002} and is given by 
\begin{eqnarray}
\label{e9}
&&\frac{f_{DH}(\phi_{p_+}, \phi_{p_-})~a^3}{k_B T} =- {\frac{(\kappa_D a)^{3}}{12 \pi}}  = \nonumber\\
&& ~~~~~~- {\textstyle\frac23} \sqrt{\frac{\pi}{a^3}}~\ell_B^{3/2}\Big({N_{p^{+}}}\phi_{p^+}+{N_{p^-}}\phi_{p^-}\Big)^{3/2},
\end{eqnarray}
where again $f_{DH}$ is the free energy density, and the inverse Debye screening length, $\kappa_D^{-1}$, is given by $\kappa_D^2 = 4\pi \ell_B \left( {c_{p^+}{N_{p^+}}^2} + {c_{p^-}{N_{p^-}}^2}\right)$, where $\ell_B$ is the Bjerrum length, $\ell_B=e^2/(4\pi \varepsilon k_B T)$, $\varepsilon = \epsilon\epsilon_0$ with $\epsilon$ the relative dielectric permittivity and $e$ the elementary charge. The assumption in the above formula is that the valency of polyions coincides with the number of monomers. 

Often the DH correlation free energy is taken in the form corresponding to a finite charge radius (see below) that of course reduces to the above expression for vanishing ionic size \cite{FALKENHAGEN19711,McQuarrie,muthu2002}. {The DH  correlation free energy is generally obtained from the thermal fluctuations around the mean-field (saddle-point) PB theory \cite{Perspective2013}, either by a coupling constant integration of the electrostatic potential of a single ion \cite{Levin_2002}, or by integrating out the quadratic electrostatic potential fluctuations around the PB saddle-point \cite{FUNINT}. The two methods are completely equivalent and either way, the inverse Debye screening length can be obtained as \cite{AVNI201970} }
\begin{eqnarray}
\kappa_D^2 = \varepsilon^{-1} \frac{\partial \rho(\psi)}{\partial \psi}{\Big\vert}_{\psi = 0} = \frac{ {\cal C}}{\varepsilon},
\label{bgcfhjsk}
\end{eqnarray}
where $\rho$ is the mobile charge density, $\psi$ the mean-field PB electrostatic potential, and ${\cal C}$ is the {\sl capacitance density} at thermal equilibrium. The Debye screening length can be interpreted as the thickness of an equivalent parallel plate condenser whose surface charges result from an  imposed potential difference. By the general statistical mechanical relationship between fluctuations and response functions, it can be expressed also in terms of the thermal fluctuations of the electrostatic potential by the Einstein formula \cite{Einstein}. 

The Debye screening length is of course straightforward to calculate in the Debye-H\" uckel framework, where the entropy of the ions is given by the ideal gas expression, and - as we will see shortly - the Legendre transform of the free energy density and its second derivatives can be calculated explicitly and analytically. It is less clear how to approach the screening problem for a general free energy which is not in the form of the ideal gas entropy, as is usually the case for the VO type theories.

The correlation free energy Eq. \ref{e9} thus quantifies the electrostatic potential fluctuations in the solution, whatever its  composition and whatever its other degrees of freedom are. The importance of electrostatic potential fluctuations is the fundamental insight of the VO theory. In some sense it can be seen also as a special case of the van der Waals theory, that recognizes the importance of the electrodynamic fluctuations  {\sl via} the dispersion interaction \cite{parsegian_2005}. Similarly to the van der Waals theory, the VO theory displays features of a mean-field theory \cite{Srivastava} even if the attractive Debye-H\" uckel term corresponds to fluctuations around a zero potential mean-field state \cite{Perspective2013}.  

The characterizing features of the theory can be and were criticized on different levels and as a consequence it has been generalized/ammended to include either better approximations, better models or both (see the discussion in the recent excellent review Ref. \cite{Sing2020}), {with some of the major new directions emerging from the implementations of the polymer field theory \cite{Pablo, Fred}, with more detailed consideration of connectivity and excluded volume effects, the scaling or ‘blob’ theory approaches \cite{Rubin}, based on the competition of thermal and electrostatic degrees of freedom that set the key length scales, as well as explicit inclusion of counterion condensation and/or localization mechanisms that includes additional degrees of freedom such as non-ionic interactions in the description of the polyelectrolyte complexation \cite{Liu}. Nevertheless, the most widely used point of departure for further developments remains the Voorn-Overbeek theory \cite{Larson2016} that endures also as the baseline for comparison \cite{Sing2020}.}

Here, however, our ambition is more technical in nature and can be formulated as follows: since the VO free energy is {\sl not} of the ideal gas form, as assumed in the DH theory that is  consistent with the Debye screening length, what is then the correct form of the screening length that is consistent with the VO or related theories? {The methodology employed is based on standard properties of the Legendre transform as reviewed by Zia et al. \cite{Zia} as well as a generalization of the screening length for a general free energy {\sl Ansatz} as developed in Ref. \cite{maggs2016}. We will write down the general equations for the screening length, based on the local curvature properties of the Legendre transform, apply them specifically to the VO model free energies and finally comment on when the screening actually reduces to the simplified and standardly assumed Debye form in the context of the polyelectrolyte condensation. The approach advocated here is rather similar in essence to the mean-field analysis of the steric \cite{Borukhov1,maggs2016}, correlation \cite{Santangelo,Bazant} and structural \cite{Blossey,Ciach} effects that result in modified Poisson-Boltzmann equations with different types of local nonlinearities which cover steric effects, higher-order derivative terms that describe the correlation effects or both for the structural effects. While these approaches certainly address the main features of the phenomena and are computationally tame, more sophisticated theories are needed to address the combined effects of packing constraints, correlation interaction and solvent structure \cite{Tarazona, Kjellander2020}. }

Our aim is thus not to generalize or improve on the VO theory. What we set out to do has a more modest but nevertheless a fundamental goal: we will address the problem of the form of the screening length, avoiding the standard definition of the Debye screening length, and making it consistent with the other terms in the VO free energy {\sl Ansatz}.

\section{General definition of screening length}

\begin{figure}
\begin{center}
  \includegraphics[scale=0.18]{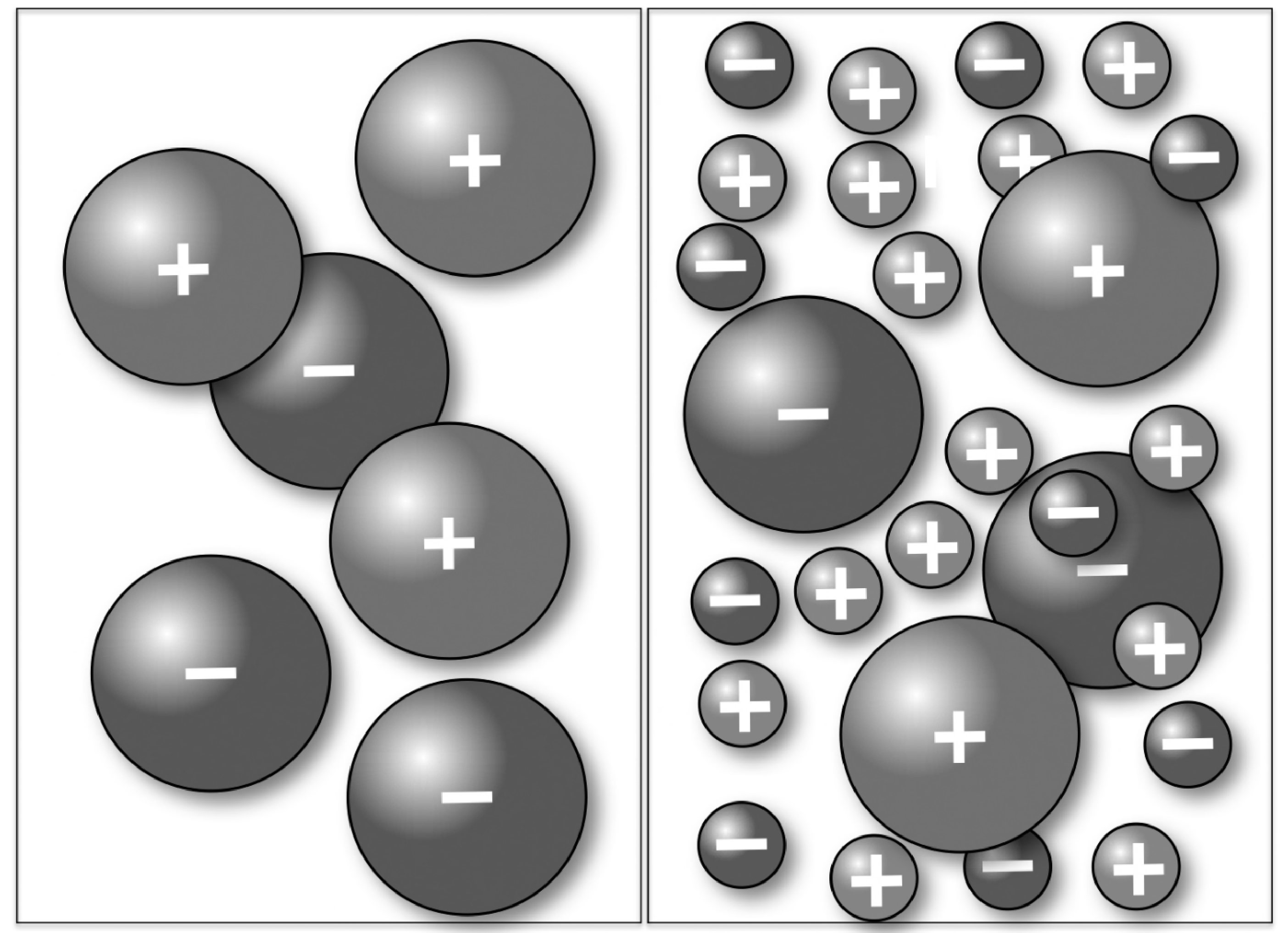}
  \caption{\label{fig:sch} A schematic representation of the two models analyzed: (left) model 1, a solution of polyions composed of different numbers of charged monomers and (right) model 2, a solution of polyions and simple monovalent salt in an aqueous solvent in both cases. The large particles are charged polymers of $N_+$ and $N_-$ monomers treated as Flory-Huggins particles.}
\end{center}  
\end{figure}

We start with the free energy, or rather free energy density, of an uncharged system composed of ${\cal N}$ components: $f(c_j) = f(c_1, c_2, \dots c_{\cal N})$. In an analysis proceeding from the Legendre transform and the local thermodynamics approximation, Maggs and Podgornik \cite{maggs2016} have recently shown that the thermodynamic potential of a charged system, where each component $j$ has a charge $e_j N_j$, can be written as 
\begin{equation}
{\cal F} [\mu_j, \psi] = - \int_V d^3{\bf r}
  \Big( {\textstyle\frac12} {\varepsilon} (\bnabla\psi)^2 + p(\mu_j + e_j
    N_j \psi)\Big),
  \label{equ3}
\end{equation}

where $p(\mu_j) = p(\mu_1, \mu_2,  
\dots \mu_{\cal N})$ is the thermodynamic pressure, or the equation of state, defined as the Legendre transform of the free energy  \cite{Widom}, $ f(c_j) - \sum_j \mu_j  c_j  = - p(\mu_j)$, while $\mu_j$ is the chemical potential of the $j$-th species. $\psi({\bf r})$ is the Legendre multiplier field, identified as the electrostatic potential, that ensures the local imposition of the Gauss' law. {Clearly the above result reduces to the standard PB form since for a two component uni-univalent electrolyte the equation of state has the ideal gas form $p(\mu \pm e \psi) = \rho_+ + \rho_- = 2 e^{\mu} \cosh{e \psi}$ \cite{Safinya}.}

While the whole derivation proceeded entirely on the mean-field level, it can be extended to the case when the Coulomb interactions are included exactly and the mean potential becomes the fluctuating local potential in a functional integral representation of the partition function as derived by Wiegel~\cite{Wiegel}. 

{In fact the field theory representation of the partition function is formally obtained with a field action at imaginary values of the electrostatic potential,  ${\cal F} [\psi] \longrightarrow {\cal F} [i\psi]$ (for details see Ref. \cite{Tomer}). The mean field or the PB solution then corresponds to the saddle-point of the field action with $\psi \longrightarrow \psi_{0}$ and free energy $F_{PB}(\psi_{0})$, while the Gaussian fluctuations around the mean field can be integrated out yielding the one-loop fluctuation corrected free energy density,} {$F = F_{PB}(\psi_{0}) + F_c$, where the correlation free energy is given by} 
\begin{eqnarray}
F_c = - {\textstyle\frac12} k_BT ~{\rm Tr} \ln{\frac{\delta^2 {\cal F}[i\psi]}{\delta\psi({\bf r})\delta\psi({\bf r}')}}{\Big\vert}_{\psi_0}.
\label{bdfhja}
\end{eqnarray}
with field Hessian 
\begin{eqnarray}
\frac{\delta^2 {\cal F}[i\psi]}{\delta\psi({\bf r})\delta\psi({\bf r}')} = \left( -\varepsilon\nabla^2 + \frac{\partial^2 p(\mu_j + i e
    N_j \psi)}{\partial \psi^2}{\Big\vert}_{\psi_0}\right). 
    ~
\end{eqnarray}
If the bulk system is electroneutral then the second derivative of $p(\mu_j + i e N_j \psi)$ is evaluated at $\psi_0 = 0$, otherwise it has to be evaluated at the value of the Donnan potential, $\psi_0 = \psi_D$.  Evaluating the last term Eq. \ref{bdfhja} in the Fourier space, subtracting the single ion contribution and assuming a vanishing size of ions, yields  after some rearrangements the first identity in Eq. \ref{e9} with the inverse screening length defined as
 \begin{equation}
 \kappa^2 \equiv  \frac{\partial^2 p(\mu_j + i e N_j \psi)}{\varepsilon ~\partial \psi^2}{\Big\vert}_{\psi_0 = 0} \!\!\!= \frac{e^2}{\varepsilon} \sum_{j,k} N_j N_k  \frac{\partial^2 p(\mu_j)}{\partial \mu_j \partial \mu_k}, 
 \label{kappadef}
 \end{equation}
where we assumed an electroneutral bulk system with the potential $\psi_0 = 0$. Invoking furthermore the Gibbs-Duhem relation, $c_{j} = {\partial p(\mu_j)}/{\partial \mu_{j}}$, we can rewrite the inverse screening length with the density derivatives, which are of course the relevant response functions. Equation \ref{kappadef} is then the proper generalization of the Debye definition of the screening length Eq. \ref{bgcfhjsk}.

{While the point charge form of the DH correlation free energy Eq. \ref{e9} is
commonly used in the VO theory, it can be straightforwardly generalized to include the finite size of the ions. In this case the electrostatic {\sl correlation free energy change} can be written as 
\begin{eqnarray}
\Delta F_c = - {\textstyle\frac12} k_BT~{\rm Tr} \ln{\frac{- \nabla^2 + \kappa^2}{- \nabla^2}}
\label{subtract}
\end{eqnarray}
with $\kappa^2$ given by Eq. \ref{kappadef}. This can be furthermore rewritten as a coupling constant integral of the resolvent operator
\begin{eqnarray}
\Delta F_c = - {\textstyle\frac12} k_B T~{\rm Tr} \int_0^1 d\lambda ~{\cal G}_{\lambda} ({\bf r}, {\bf r}')
\label{ccintegral}
\end{eqnarray}
with the Green's function ${\cal G}_{\lambda} ({\bf r}, {\bf r}')$ given as a solution of 
\begin{eqnarray}
\left( - \nabla^2 + \lambda \kappa^2\right) {\cal G}_{\lambda} ({\bf r}, {\bf r}') = \delta({\bf r} - {\bf r}')
\end{eqnarray}
where one needs to subtract the contribution of the $\kappa^2 = 0$ solution, see Eq. \ref{subtract}. This can be obtained straightforwardly as \cite{FALKENHAGEN19711,McQuarrie,Levin_2002}
\begin{eqnarray}
{\cal G}_{\lambda} ({\bf r}, {\bf r}) = \frac{\lambda^{1/2}\kappa}{4\pi ~(1 + \lambda^{1/2}\kappa a)}
\end{eqnarray}
since in the evaluation of $\rm Tr$ only the value at identical arguments is needed. Insering back into Eq. \ref{ccintegral} one derives the expression commonly used in the free energy {\sl Ansatz} of most current VO type theories \cite{Sing2020,Wang2021,Muthu2021}
\begin{eqnarray}
\Delta F_c = - \frac{V}{4\pi a^3} \left( \ln{(1 + \kappa a)} - (\kappa a) + {\textstyle\frac12} (\kappa a)^2 \right),
\end{eqnarray}
with ${\rm Tr} "1" = V$, the volume of the system. 
In the limit $a \longrightarrow 0$ this reduces back to Eq. \ref{e9}. Notably in the above derivation the Debye charging process \cite{FALKENHAGEN19711,McQuarrie,Levin_2002} is recognized simply as the "coupling constant integration" of the resolvent operator.}

The form of the inverse Debye screening length as given in Eq. \ref{kappadef} together with the curvature duality of the Legendre transform \cite{Zia}, leads to its straightforward and elegant calculation for any form of the free energy even when the equation of state $p(\mu_j)$ is not explicitly available \cite{maggs2016}. 

We can test these expressions on the original two component Debye-H\" uckel theory with $N_1=N_2=1$ and volume fractions $\phi_{1,2} = a^3 c_{1,2}$, where the entropy of the uncharged systems is assumed to be just the ideal gas entropy 
\begin{align}
  &\frac{f_{D}(\phi_1, \phi_2)~a^3}{k_B T} = {\phi_1} \left(\log{\phi_1}-1\right) + {\phi_2} \left(\log{\phi_2} -1 \right), 
  \label{bcrtywo1}
\end{align}
the Legendre transform of which can be obtained analytically in the simple form of 
\begin{equation}
p_D(\mu_1, \mu_2) = \frac{k_BT}{a^3} \left( e^{\beta \mu_1} + e^{\beta \mu_2} \right) = p(\mu_1) + p(\mu_2), 
  \label{dapois2}
\end{equation}
clearly additive in the two components. The inverse Debye length from Eq. \ref{kappadef}, or equivalently from the expansion of the PB equation, then follows as
\begin{equation}
 \kappa_D^2 =  \frac{4\pi \ell_B}{a^3} \left( N_1 \phi_1 + N_2 \phi_2\right) = 4\pi \ell_B ( N_1^2 c_1 + N_2^2 c_2),
  \label{negiouv2}
\end{equation}
where $\ell_B$ is again the Bjerrum length. This is of course nothing but the standard Debye screening length. 

In what follows we will analyze the screening length of two models, see Fig. \ref{fig:sch}: a model of polyions (model 1) that reduces to the Flory–Huggins theory for uncharged polymers and a model of polyions (model 2) with monovalent salt that is a combination of the Flory-Huggins theory for polyions and  DH  theory for simple salt ions \cite{muthu2010,Wang2021}. The ensuing screening length will in general differ substantially from a simple Debye screening form as we elucidate next.

\section{Screening in Voorn-Overbeek type theories}

\begin{figure}
\begin{center}
\centerline{
  \includegraphics[scale=0.4]{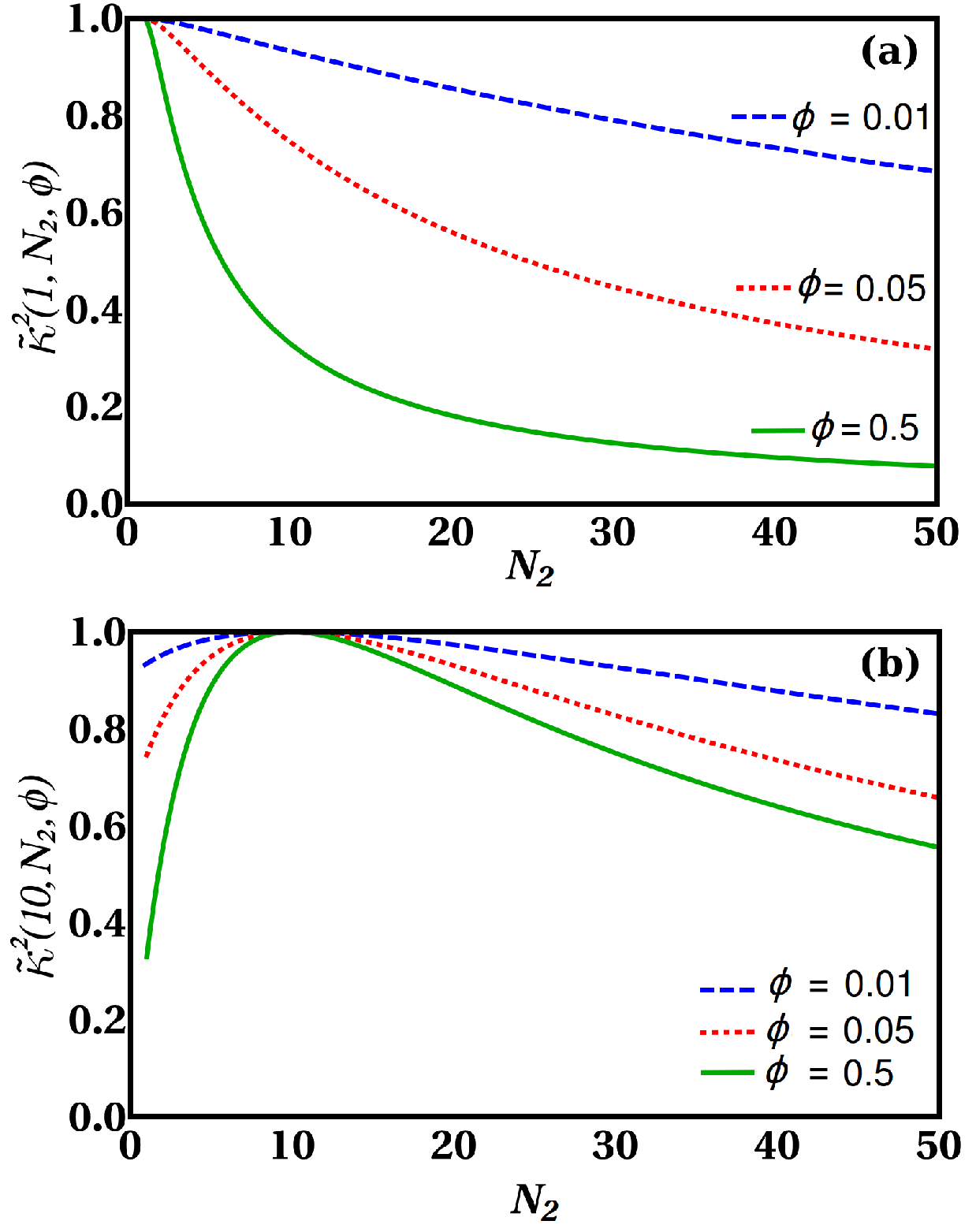}}
  \caption{ The dependence of the ratio  $\tilde\kappa^2(N_1, N_2, \phi_M)$, Eq. \ref{ratio1},  on $N_2$ for two different values of $N_1 = 1, 10$. In the first, highly asymmetric case ($N_1 = 1$), the screening length monotonically increases, whereas in the second case ($N_1 = 10$) it shows non-monotonic behavior. This behavior centered around the symmetric state, $N_1 = N_2$, obviously depends on the nature of asymmetry in the system. \label{Fig1}}
\end{center}  
\end{figure}

{We analyze a couple of non-trivial generalizations of the Debye screening for models that are standardly used in the description of complexation in mixtures of different polyelectrolytes \cite{WangMM2018,Sing2020,Zheng2021,muthu2010,Larson2016,Wang2021}.}

{\sl Model 1}. We start with a polydisperse polyion mixture, {\sl model 1 }, composed of species ``1" at concentration $c_1$, itself composed of $N_1$ monomers, each of charge $e$, and species ``2" at concentration $c_2$, itself composed of $N_2$ monomers, each of charge $-e$, in an aqueous solvent of (water) molecules of diameter $a$. This model has been used for the description of complexation in mixtures of oppositely charged polyelectrolytes \cite{WangMM2018,Sing2020,Zheng2021}.

Expressed in terms of the volume fractions $\phi_1, \phi_2$, defined as $ \phi_{1,2} = a^3 c_{1,2} N_{1,2}$, the Flory-Huggins lattice level free energy density is \cite{Teraoka, muthu2010}
\begin{align}
&&\frac{f(\phi_1, \phi_2)~a^3}{k_B T} = \frac{\phi_1}{N_1} \log{\phi_1} + \frac{\phi_2}{N_2} \log{\phi_2} \nonumber\\
&& + (1 - \phi_1-\phi_2)\log{(1 -\phi_1-\phi_2)}.
  \label{bcrtywo}
\end{align}
We omit the $\chi$ interaction term as it is irrelevant for subsequent derivations. The Hessian of the pressure $p(\mu_1, \mu_2)$ is positive definite, and according to Eq. \ref{kappadef}, the inverse Debye screening length is given by
\begin{equation}
 \kappa^2 =   \frac{e^2}{\varepsilon} \left( N_1^2\frac{\partial^2 p}{\partial \mu_{1}^2} - 2 N_1N_2\frac{\partial^2 p}{\partial \mu_{1}\mu_{2}} + N_2^2\frac{\partial^2 p}{\partial \mu_{2}^2}\right).
  \label{negiouv}
\end{equation}
Contrary to the original two component Debye theory, derived in the previous section, the chemical potential derivatives cannot be evaluated explicitly as the lattice entropy does not posses a simple analytical Legendre transform.  

At this point, however, one can take recourse to the general properties of the Legendre transform, specifically to its curvature duality property. One of the fundamental properties of the Legendre transform is that the Hessian of the Legendre transform is the inverse of the Hessian of the function itself \cite{Zia}, so that one can write
\begin{equation}
  \sum_{m=1}^{2} \frac{\partial^2 p(\mu_1, \mu_2)}{\partial \mu_{j} \partial \mu_{m}} \frac{\partial^2 f(c_1, c_2)}{\partial c_{m} \partial c_{k}} = \delta_{jk}
  \label{invgyui}
\end{equation}
where all the matrices are $2 \times 2$. This implies that the local curvature of the Legendre transform is inverse to the local curvature of the original function in a manner reminiscent of the uncertainty relation, as observed by Zia {\sl et al.} \cite{Zia}. The above relation remains valid only for strictly convex functions so that neither derivative ever vanishes. 

The matrix of partial derivatives $\frac{\partial^2 f(c_1, c_2)}{\partial c_{m} \partial c_{k}}$ from the free energy Eq. \ref{bcrtywo} can be calculated straightforwardly and from Eq. \ref{invgyui} one then obtains the derivatives $\frac{\partial^2 p(\mu_1, \mu_2)}{\partial \mu_{j} \partial \mu_{m}} $ which then yield $\kappa^2$ from the combination in Eq. \ref{negiouv}, so that
\begin{eqnarray}
 &&\kappa^2(N_1, N_2, \phi_1, \phi_2) \nonumber\\
 &&~~~~~~=\frac{4\pi \ell_B}{a^3}  \frac{u  \left(N_1 \phi _1+N_2 \phi _2\right)+4 N_1 N_2 \phi _1 \phi _2}{ \left(N_1 \phi _1+N_2 \phi _2+ u \right)},
 \label{kap1}
\end{eqnarray}
where we used the abbreviation $u = (1 -\phi_1-\phi_2)$ and the Bjerrum length $\ell_B$ was defined before.  

In general the inverse square of the screening length is therefore {\em not} a linear function of the concentrations as is the case for the Debye screening length. The bulk electroneutrality furthermore restricts the concentrations of components to  $N_1c_1- N_2c_2 = 0$ or equivalently 
\begin{eqnarray}
\phi_1 - \phi_2 = 0,
\end{eqnarray}
while both $N_1, N_2$ can still remain arbitrary.  Denoting $\phi_{1,2}=\phi_M$,  Eq.~\ref{kap1} can be recast in the form
\begin{widetext}
\begin{eqnarray}
\kappa^2(N_1, N_2, \phi_M)=  \frac{4\pi \ell_B}{a^3} \frac{ \left(N_1+ N_2 \right) \phi_M + 2 \big(2 N_1N_2 - (N_1 + N_2)\big) \phi_M ^2 }{ \big(1 +({N_1} + {N_2} -2) \phi_M\big)}.
\end{eqnarray}
\end{widetext}
In the case of $N_1 = N_2 = 1$, as well as for any symmetric case, $N_1 = N_2$, this obviously reverts back to the standard Debye form . However, for any other case the screening length is a much more complicated function of the volume fractions, or concentrations of the species.

On Fig. \ref{Fig1} we show the dependence of the ratio 
\begin{equation}
    \tilde\kappa^2(N_1, N_2, \phi_M) = \frac{\kappa^2(N_1, N_2, \phi_M)}{(N_1+N_2)\phi_M}
    \label{ratio1}
\end{equation} 
for an electroneutral system. The denominator would be the standard inverse square of the Debye screening length expected for point ions. Clearly the dependence of $\tilde\kappa^2$ is in general not a linear function of $N_1, N_2$ and can in addition show strong non-monotonic behavior with a minimum at the standard Debye screening value corresponding to $\tilde\kappa^2 = 1$. This non-monotonic regime in the vicinity of $N_1 = N_2$ state, at which the screening length exhibits a minimum value, is the more pronounced the larger is the volume fraction of the polyions. {Obviously the non-symmetric systems are then under-screened with the screening length exceeding the Debye screening length expectations.}

{\sl Model 2}. We now proceed to a more complicated system, {\sl model 2}, composed of a uni-univalent salt as well as a polydisperse polyion mixture that we already discussed above. {We refer to this model as the Voorn-Overbeek-type model, which has been recently used to model complexation in mixtures of oppositely charged polyelectrolytes with explicit salt ions \cite{muthu2010,Larson2016,Wang2021}.}

Positively ``3" and negatively ``4" charged polymers have $N_3$ and $N_4$ monomers, while the salt is composed of  univalent positively ``1" and negatively ``2" charged simple salt ions. The free energy can be taken in a form generalizing the Flory-Huggins lattice entropy as 
\begin{widetext}
\begin{eqnarray}\nonumber
  \frac{f(\phi_1, \phi_2, \phi_3, \phi_4)~a^3}{k_B T} &=& {\phi_1} \log{\phi_1}  +  {\phi_2} \log{\phi_2} + \frac{\phi_3}{N_3} \log{\phi_3} + \frac{\phi_4}{N_4} \log{\phi_4} \\
  &+& (1 - \phi_1-\phi_2-\phi_3 - \phi_4)\log{(1 - \phi_1-\phi_2-\phi_3 - \phi_4)}.
  \label{VO1}
\end{eqnarray}
\end{widetext}
The first two terms describe the simple monovalent salt, the next two terms correspond to polyions, while the last term is the solvent entropy. The above free energy is clearly a straightforward generalization of Eq. \ref{bcrtywo}.

We now proceed in the same way as before, except that now the analysis is a bit more involved since we have a four component system: salt and polyions, so that both the free energy, $f(\phi_1, \phi_2, \phi_3, \phi_4)$, as well as the pressure, $p(\mu_1, \mu_2, \mu_3, \mu_4)$, are functions of four variables and consequently the Hessian matrices of derivatives will now be $4 \times 4$, instead of $2 \times 2$, with the algebra correspondingly more difficult, but not unmanageable analytically. 

The Debye length in this case is obtained with complete analogy to Eq. \ref{negiouv} as
\begin{equation}
\kappa^2 =  \frac{e^2}{\varepsilon} \left(  {\partial_{\mu_1}}\!\!-\!\!{\partial_{\mu_2}}\!\!+\!\!N_3 {\partial_{\mu_3}}\!\!-\!\!N_4 {\partial_{\mu_4}} \right)^2 p(\mu_1, \mu_2, \mu_3, \mu_4).
  \label{negiouv11}
\end{equation}
In order to evaluate this, we need to invert the Hessian of the original function and calculate the Hessian of its Legendre transform through the curvature duality relation of the Legendre transform, Eq. \ref{invgyui}, which in this case assumes the following form 

\begin{eqnarray}
&&\sum_{m=1}^{4} \frac{\partial^2 p(\mu_1, \mu_2, \mu_3, \mu_4)}{\partial \mu_{j} \partial \mu_{m}} \frac{\partial^2 f(c_1, c_2, c_3, c_4)}{\partial c_{m} \partial c_{k}} =  \nonumber\\
&& ~~~~~~= \sum_m p_{j,m} f_{m,k} = \delta_{jk},
  \label{invgyui2}
\end{eqnarray}
where we have introduced the following notation for the double derivatives of free energy density and pressure
\begin{equation}
    f_{m,k} = \partial_{c_{m}} \partial_{c_{k}} f(c_1, c_2, c_3, c_4)
\end{equation}
with $ f(\phi_1, \phi_2, \phi_3, \phi_4)$ given by Eq. \ref{VO1}, and 
\begin{equation}
    p_{m,k} = \partial_{\mu_{m}} \partial_{\mu_{k}} p(\mu_1, \mu_2, \mu_3, \mu_4)
\end{equation}
Again, Eq. \ref{invgyui2} implies that the local curvatures of the Legendre transforms are inverse to each other as observed before \cite{Zia}. 

\begin{figure}
  \includegraphics[scale=0.4]{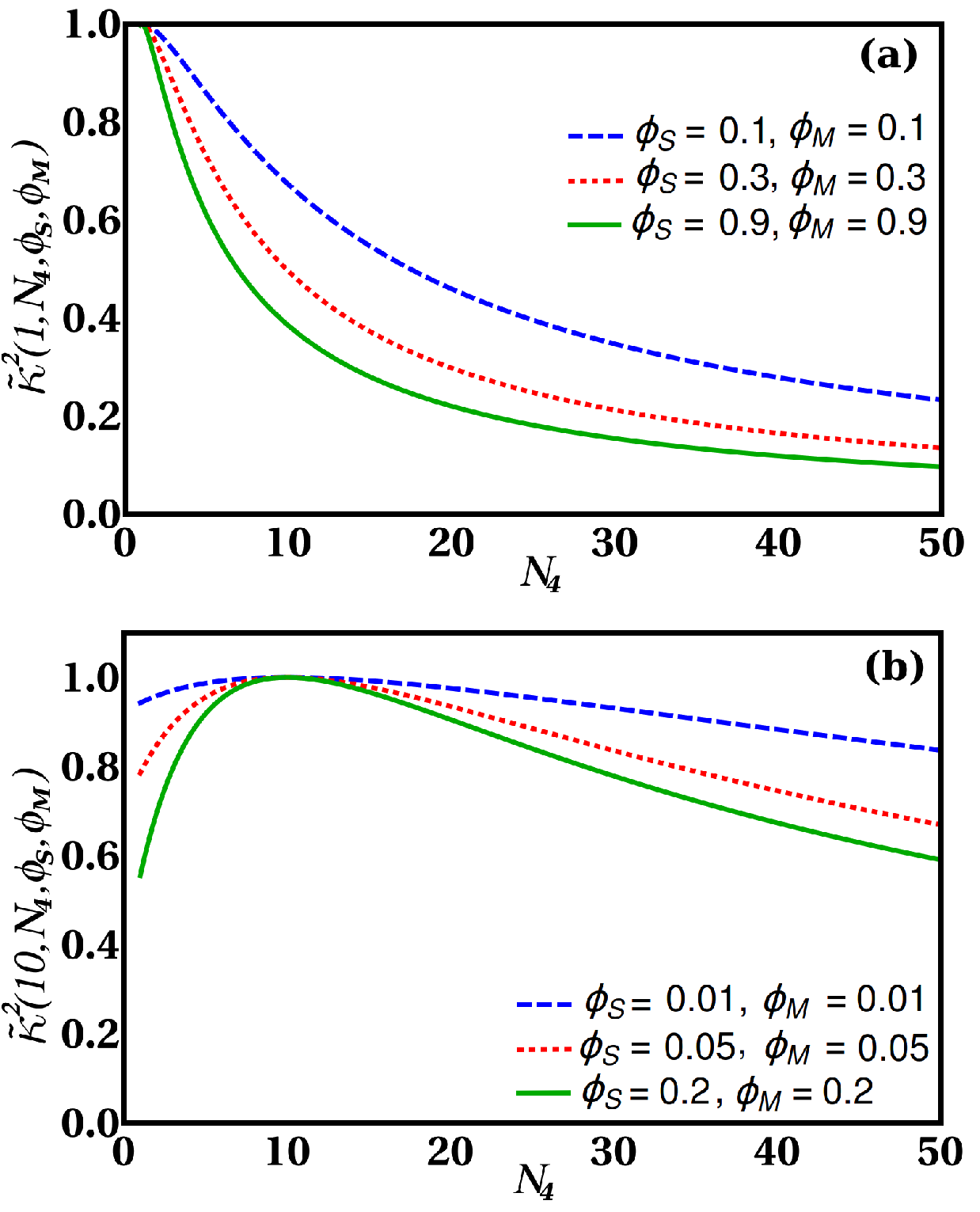}
  \caption{Dependence of the ratio $\tilde\kappa^2(N_3, N_4, \phi_s, \phi_M)$, Eq. \ref{ratio2}, on $N_4$ for two different values of $N_3 = 1, 10$ and $\phi_s, \phi_M$ as indicated in the figure. The screening length attains a minimum for the symmetric case, $N_3 = N_4$, with the value at the minimum depending on the two volume fractions. Again, $\tilde\kappa^2(N_3, N_4, \phi_s, \phi_M) = 1$ corresponds to the simple Debye form of the screening length given by $\kappa^2(N_3, N_4, \phi_s, \phi_M) = 2 \phi_s + (N_3 + N_4) \phi_M$. \label{Fig3}}
\end{figure}

\begin{figure}
  \includegraphics[scale=0.4]{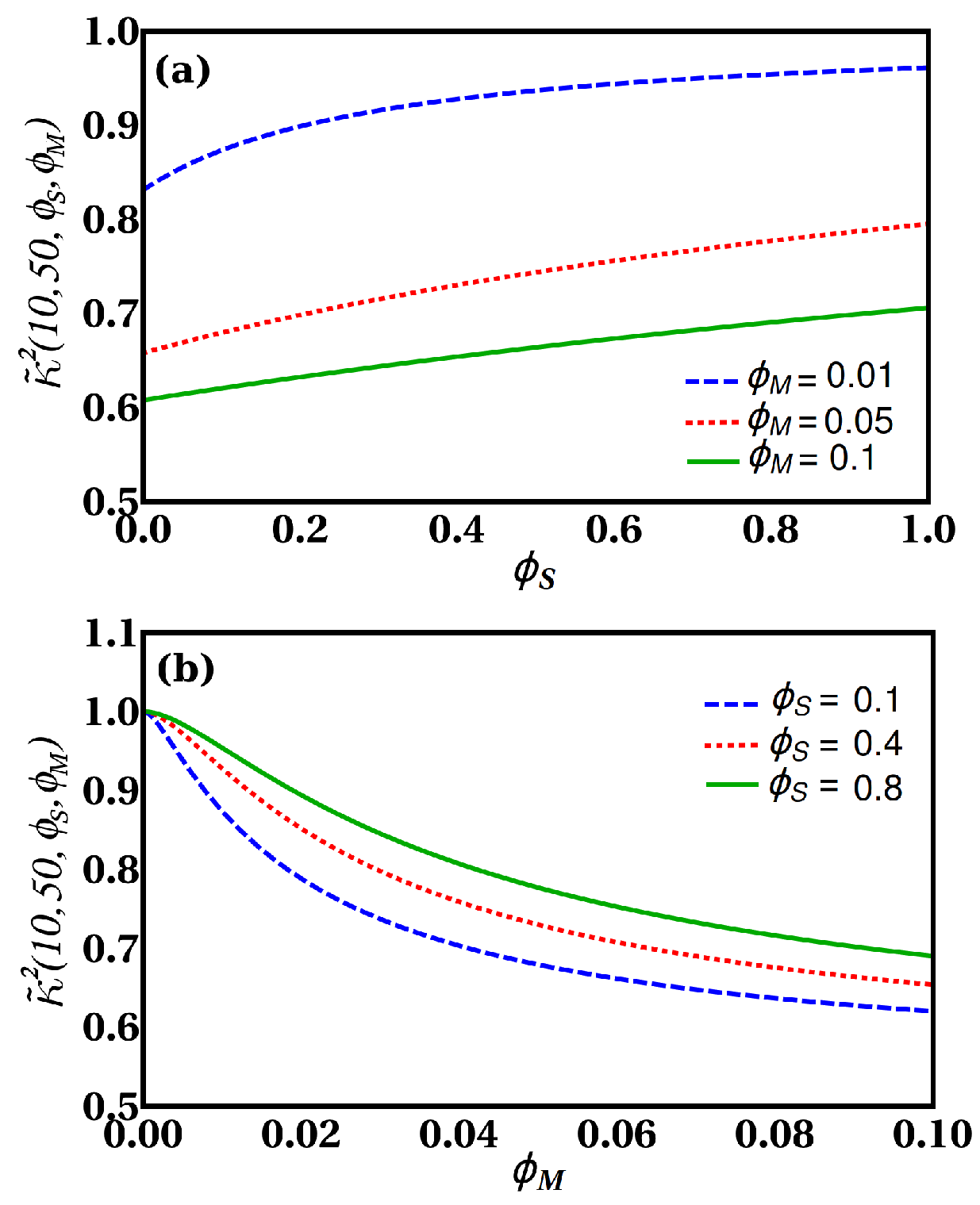}
  \caption{Dependence of the ratio $\tilde\kappa^2(N_3, N_4, \phi_s, \phi_M)$ on  $\phi_s, \phi_M$, Eq. \ref{ratio2},  for $N_3$ = 10 and $N_4$ = 50 as indicated in the figure . The symmetric case of $N_3 = N_4$ reduces back to the standard Debye length, coinciding with $\tilde\kappa^2(N_3, N_4, \phi_s, \phi_M) = 1$ on the figure. The dependence of the screening length on the two volume fractions deviates significantly from the Debye screening length for non-vanishing volume fractions $\phi_s, \phi_M$, and shows opposite trends as a function of $\phi_s$ and $\phi_M$, {\sl i.e.}, one is increasing and the other one is decreasing. \label{Fig4}}
\end{figure}
Expressing the derivatives of the equation of state   $\partial_{\mu_{j}} \partial_{\mu_{m}} p(\mu_1, \mu_2, \mu_3, \mu_4)$ with the derivatives of the free energy $ \partial_{c_{m}} \partial_{c_{k}} f(c_1, c_2, c_3, c_4)$ and inserting the result into Eq. \ref{negiouv11}, we can then obtain the inverse square of the screening length in terms of the derivatives of the equation of state in the form analogous to Eq. \ref{negiouv} as 
\begin{widetext}
\begin{eqnarray}
\kappa^2 = \frac{e^2}{\varepsilon}\Big(p_{1,1}- 2 p_{1,2} +  p_{2,2} + 2 N_3 p_{1,3}- 2 N_3 p_{2,3} - 2 N_4 p_{1,4} + 2 N_4 p_{2,4} + N_3^2 p_{3,3} - 2N_3N_4 p_{3,4} 
 &+& N_4^2 p_{4,4}\Big).
\label{negiouv3}
\end{eqnarray}
\end{widetext}
Evaluating the matrix inversion of $f_{m,k}$ explicitly from Eq. \ref{invgyui2} we furthermore obtain

\begin{widetext}
\begin{eqnarray}\nonumber
\kappa^2 &=& \frac{4\pi \ell_B}{a^3~{\cal D}et} \Bigg(\phi_1 \Big(1-\phi_2  +(N_3-1) \phi_3+(N_4-1) \phi_4\Big) + \phi_2\Big(1-\phi_1  +(N_3-1) \phi_3+(N_4-1) \phi_4\Big)\\\nonumber
&+&\Big(2 \phi_1\phi_2 - 2N_3\phi_1\phi_3 + 2N_3\phi_2\phi_3 + 2N_4\phi_1\phi_4 - 2N_4\phi_2\phi_4 +2\phi_3\phi_4\Big)+N_3\phi_3 \Big(1-\phi_3 +(N_4-1) \phi _4\Big) \\
&+& N_4\phi_4\Big(1  -\phi_4 + (N_3-1) \phi _3\Big)\Bigg) 
\label{ka}
\end{eqnarray}
\end{widetext}

where ${\cal D}et$ is given by $${\cal D}et = \big(1  + (N_3-1) \phi_3 +(N_4-1) \phi_4\big).$$ As in the case of {\sl Model 1}  in general the inverse square of the screening length is {\em not} a linear function of the concentrations. 

In order to investigate the case of a bulk electroneutral system we need to evaluate the above expression with electroneutrality condition $c_1 - c_2 + N_3c_3 - N_4 c_4 = 0$ or equivalently
\begin{eqnarray}
\phi_1 - \phi_2 + \phi_3 - \phi_4 = 0.
\end{eqnarray}
The electroneutrality condition, involving now four variables, exerts less of a constraint on the values of the different volume fractions as in the case of {\sl Model 1}, where we had a system with only two components. 

Let us first consider the limit of $\phi_1 = \phi_2 = \phi_s$ and $\phi_3 = \phi_4 = \phi_M$, {\sl i.e.}, the salt and the polyions are electroneutral separately, being just a particular case of the general electroneutrality condition. 

In this case we obtain for the inverse square of the screening length the expression
\begin{eqnarray}\nonumber
\kappa^2 &=& \frac{4\pi \ell_B}{a^3~{\cal D}et} \Bigg(2 \phi_s \Big( 1  +(N_3\!\!+\!\!N_4\!\!-\!\!2) \phi_M \Big)  +2N_3N_4\phi_M^2 \nonumber\\
&+& N_3\phi_M \Big(1 +(N_4\!\!-\!\!2) \phi _M\Big)\!\!+\!\! N_4\phi_M\Big(1 + (N_3\!\!-\!\!2) \phi _M\Big)\Bigg)~\nonumber\\
~
\label{kap}
\end{eqnarray}
with 
${\cal D}et = 1  + (N_3 + N_4 -2) \phi _M$. The screening length in this case is obviously much more complicated then in the case of the Debye screening length and is partitioned jointly between the simple salt and the polyions. 

On Fig. \ref{Fig3} we show the dependence of the screening ratio 
\begin{equation}
    \tilde\kappa^2(N_3, N_4, \phi_s, \phi_M) = \frac{\kappa^2(N_3, N_4, \phi_s, \phi_M)}{(2 \phi_s + (N_3 + N_4) \phi_M)}
    \label{ratio2}
\end{equation}
for an electroneutral system. The denominator is again the expected ``naive" Debye screening length. The dependence of the screening length is strongly non-monotonic with a minimum at the ``naive" Debye value for $N_1 = N_2$ and exhibits pronounced under-screening with $ \tilde\kappa^2 \leq 1$. The under-screening depends on the number of monomers the broader is the regime displaying this non-monotonicity.

Figure~\ref{Fig4} is the same as Fig.~\ref{Fig3} except that we show the dependence of the screening ratio  as a function $\phi_s$ and $\phi_M$. The dependence of the screening ratio on both $\phi_s$ and $\phi_M$ is monotonic, but is an increasing function in the former and a decreasing function in the latter case, but nevertheless exhibiting under-screening in both cases. This behavior remains valid for different values of the number of charged monomers of the polyions, $N_3$ and $N_4$. The dependence on $\phi_M$ furthermore indicates that the ``naive" Debye screening length represents a minimal screening length, and the the system is strongly overscreened.
\section{Discussion and conclusions}

{The Debye-H\" uckel electrostatic correlation (fluctuation) free energy that enters the Voorn-Overbeek type theories is the only electrostatic contribution to the free energy because the system is considered in the bulk and the homogeneity eliminates any other mean-field electrostastic contribution(s). The fluctuational part of electrostatics, usually calculated in some variant of the one-loop approximation, depends on the screening length which was identified as the standard Debye screening length, based on the assumption that the polyions act as point particles and that the conformational degrees of freedom do not contribute to screening, so that the simple ions as well as the polyions contribute proportionately to the screening.  The major assumptions inherent in this approach are the point particle approximation for the polyions and omission of the conformational fluctuations of the polyions, both of which have been discussed and generalized by many later developments (for a recent review see \cite{Sing2020}).}

{There are two separate issues with this approach: one is the expression for the screening length consistent with the other terms in the free energy, which we address in this paper, and the other one is the accuracy of the free energy {\sl Ansatz}, based on the local density approximation and the lattice gas entropy, to describe the effects of the hard core repulsions, which we do not scrutinize. In this work, our modest goal was not to introduce any new approach or model, but to remain well within the Voorn-Overbeek paradigm, {\sl except} for the nature of screening and the associated screening length, which we derive self-consistently not by assuming {\sl a priori} a standard Debye form, but by deriving it explicitly in such a way that it is consistent with all the other terms in the Voorn-Overbeek type free energy {\sl Ansatz}. This was accomplished by studying the Legendre transform of the Voorn-Overbeek free energy and based on this, calculating the Hessian of electrostatic fluctuations around the mean field as described in detail in Ref. \cite{maggs2016}. The fluctuations around the mean field lead to the standard Coulomb fluid correlation free energy, {\sl except} that the screening length in general differs from the Debye form. The difference can be characterized broadly as under-screening, $\kappa \leq \kappa_D$, and thus implies a reduces correlation contribution to the free energy relative to the Voorn-Overbeek theory. This becomes particularly relevant for non-symmetric systems where the polycations and polyanions components are composed of a different number of monomers. 
 }

{Unlike the case of the ion screening in simple electrolytes, the screening properties of the polyelectrolyte solutions, which are crucial for the onset of complexation phenomenology \cite{Tirell2016}, depend importantly on the polyion chain conformation, so that the ionic screening and the polymer chain conformations self-consistently determine the nature of screening \cite{DOBRYNIN20051049,Muthu2017}. The naive approach that would simply treat the polymers as finite size particles misses the crucial point of the connection between polymer conformations and screening, whose experimental fingerprint still remains poorly understood \cite{Tirell2018}. This connection, epitomized by Muthukumar as "double screening", couples excluded volume and electrostatic interactions in a polyelectrolyte solution \cite{Muthudouble} and has been later elaborated by different attempts, most notably the field-theoretic renormalized Gaussian fluctuation theory that accounts self-consistently for the coupling between polyelectrolyte chain conformations and electrostatic interactions in the screening cloud of the polyions \cite{Wang2017,Wang2018}.}

{While our contribution is mostly methodological, its usefulness was already demonstrated before in the case of an asymmetric lattice gas \cite{maggs2016,Kornyshev2022}, and in the case of structural interactions in ionic liquids \cite{Blossey}. We are convinced that it should have important consequences also for various generalizations of the Voorn-Overbeek theory \cite{muthu2010,Larson2016,WangMM2018,Sing2020,Zheng2021,Wang2021} that are often based on complicated expressions,  way beyond the ideal gas form, and the corresponding screening lengths cannot be assumed to be of the Debye form. We do not and cannot claim that our methodology can improve on the more fundamental approaches \cite{Fredrickson, Muthu2017, Wang2017}, but it does introduce a measure of self-consistency and a possibility for a simple and fast generalization of the electrostatic correlation free energy in the cases where screening is vastly different from the case of the simple salts.}


A general conclusion that emerges from this analysis is that the Debye form of the screening length is incompatible with any theory that is not based on an ideal gas entropy term. There are limiting cases, however, such as completely symmetric systems, where even more complicated free energies lead back to the Debye screening length.  We give explicit general formulas for the screening length that are valid for any form of the free energy, including the ideal gas free energy, the Flory-Huggins free energy or the Voorn-Overbeek free energy and find that the screening length exhibits a non-monotonic under-screening behavior as a function of the number of monomers of the polyions, and displays a different functional dependence on the volume fractions of the components then the standard Debye screening length. We believe the importance of our methodology is not only to correctly evaluate the proper screening length but also to underline the consistency one needs to strive for in defining it.

\section{Acknowledgement}
The authors would like to acknowledge the support of the 1000-Talents Program of the Chinese Foreign Experts Bureau, as well as the support of the School of physical sciences, University of the Chinese Academy of Sciences, Beijing and the Institute of the physics, Chinese Academy of Sciences, Beijing. RP would like to thank Tony Maggs for his illuminating comments on an earlier version of this manuscript. 
\bibliography{Manuscript}

\end{document}